# Physics-driven discovery and bandgap engineering of hybrid perovskites

Sheryl L. Sanchez,[1] Elham Foadian[1], Maxim Ziatdinov,[2,3] Jonghee Yang[1], Sergei V. Kalinin,[1] Yongtao Liu[*2,] and Mahshid Ahmadi[1*]

[1] *Institute for Advanced Materials and Manufacturing, Department of Materials Science and Engineering, The University of Tennessee Knoxville, Knoxville, Tennessee, 37996, United States*

[2] *Center for Nanophase Materials Sciences, Oak Ridge National Laboratory, Oak Ridge, TN 37831, USA*

[3] *Computer Science and Engineering Division, Oak Ridge National Laboratory, Oak Ridge, TN 37831, USA*

*Corresponding author emails:

mahmadi3@utk.edu

liuy3@ornl.gov

**Abstract**

The unique aspect of the hybrid perovskites is their tunability, allowing to engineer the bandgap via substitution. From application viewpoint, this allows creation of the tandem cells between perovskites and silicon, or two or more perovskites, with associated increase of efficiency beyond single-junction Schokley-Queisser limit. However, the concentration dependence of optical bandgap in the hybrid perovskite solid solutions can be non-linear and even non-monotonic, as determined by the band alignments between endmembers, presence of the defect states and Urbach tails, and phase separation. Exploring new compositions brings forth the joint problem of the discovery of the composition with the desired band gap, and establishing the physical model of the band gap concentration dependence. Here we report the development of the experimental workflow based on structured Gaussian Process (sGP) models and custom sGP (c-sGP) that allow the joint discovery of the experimental behavior and the underpinning physical model. This approach is verified with simulated data sets with known ground truth, and was found to accelerate the discovery of experimental behavior and the underlying physical model. The d/c-sGP approach utilizes a few calculated thin film bandgap data points to guide targeted explorations, minimizing the number of thin film preparations. Through iterative exploration, we demonstrate that the c-sGP algorithm that combined 5 bandgap models converges rapidly, revealing a relationship in the bandgap diagram of $MA_{1-x}GA_xPb(I_{1-x}Br_x)_3$. This approach offers a promising method for efficiently understanding the physical model of band gap concentration dependence in the binary systems, this method can also be extended to ternary or higher dimensional systems.

**Introduction**

Organic-inorganic hybrid perovskites have gained significant attention in the field of materials science due to their superior properties and ease of manufacturability. These materials have shown remarkable progress in their power conversion efficiency (PCE), with advancements from 3.8% to 25.7% for single junction and above 30% for tandem perovskite and silicon solar cells.[1-3] This has generated considerable excitement, as perovskite materials hold the potential to replace traditional materials in various optoelectronic applications. Recent research has shown that compositional engineering by mixing cations and halides can exhibit improved performance compared to other alternatives.[4-7] However, these material advancements can come at the cost of stability issues. For instance, there may be decomposition of the original precursor solutions or intrinsic halide segregation, leading to complex phase spaces.[8-11] The interplay of various components can give rise to complex behaviors across the multicomponent phase diagrams, necessitating a more nuanced understanding of their properties. Nonetheless, the tuning of bandgap and stability is of great interest in the field of optoelectronics. For instance, tandem solar cells greatly benefit from bandgap engineering to optimize the absorption of light across different wavelengths, allowing a broader spectrum of light to be captured. Similarly, light emitting diodes (LEDs) rely on different bandgaps to produce a range of colors.[12-16] Understanding these properties is essential in optimizing the performances and stability of perovskite materials across a spectrum of applications.

The interplay of the various endmembers can give rise to complex phase diagrams when looking at solid solutions with two and three phase regions. These various phases can lead the changes in bandgap evolution. For the bandgap, substitution on the A site in the perovskite structure with a chemical formula of $ABX_3$ (A = monovalent cations; B= divalent metal cations such as $Pb^{2+}$ or $Sn^{2+}$; X = halides) can directly lead to changes in bandgap. For example, incorporating an larger or oversized A cation such as formamidinium (FA; ionic radius ($r_A$) = 256)[17, 18] or dimethylammonium ($r_A$ = 272 picometers (pm))[17] compared to methylammonium (MA; $r_A$ = 217 pm)[17, 19], manifests smaller bandgaps of the perovskites.[20, 21, 22] Substitution of I into Br or Cl can also lead to a bandgap decrease.[23-25] However, some reports have shown that the size and geometry of the A cation can affect the bond length and angle between the B cation and X halide species[26], which can also modulate bandgap. In addition, substitution often also causes phase transitions and bandgap can change abruptly at the structural phase transition boundaries.[20, 27] These effects of the A, and X site substitution can give rise to complex behaviors across the multicomponent phase diagrams broadly explored towards tunability and stability optimization. The bandgap-component dependence has been controversially reported.[28-33] Which requires a better understanding of this relationship. It is also important to note that the optically determined bandgap can differ from the energy level landscape due to the potential presence of shallow defect levels, Urbach tails, and phenomena such as phase separation or chemical instability can compromise the effective bandgap.[34, 35] These effects can be often identified from the photoluminescent and adsorption behaviors such as peak shapes.[36]

$MAPbI_3$ has been widely used as the light absorber in perovskite solar cells, but it has shown to be unstable in ambient conditions due to the migration and electrochemical reaction of ions with the interface layer under operation condition[37, 38]. The movement of the ions within the perovskite layer, especially lead ions can lead to the formation of defects and traps that can degrade the performance of the solar cell. To address this issue, larger organic cations such as FA or guanidinium (GA) has been added to the MAPbI$_3$ in the MA sites, to mitigate the stability and ion migration.[39-41] This is due to the ability for a bulkier structure to provide steric hindrance making it harder for moisture to enter the crystal lattice, it can help stabilize the crystal structure preventing phase transitions and also hinder the migration of ions within the perovskite lattice. Tolerance factors have been used to forecast the stabilities of these modified perovskite structures. The tolerance factor of the MAPbI$_3$ perovskite is ~0.91, which deviates from 1, suggesting its crystal structure is distorted from the cubic structure[42]. The $r_A$ of GA (278 pm)[17] is larger than MA (217 pm)[41], which reduces crystal distortion and increases the tolerance factor, resulting in improved long-term stability.[40, 41, 43].

Synthesis methods can be broadly categorized into manual and high-throughput approaches. Manual synthesis involves hand-crafted processes, often resulting in the creation of one sample at a time. In contrast, high-throughput synthesis automates the creation of multiple samples, significantly accelerating the study of materials and their properties. Machine Learning (ML) methods offer the advantage of predicting the composition dependence of the bandgap, reducing the number of samples required. With a known physical model that predicts this composition dependence, the desired composition can be easily determined. If the concentration space is explored using high-throughput synthesis, a physical or non-parametric model can be derived from the experimental data.

Though high-throughput synthesis is quicker than manual methods, which can be time-consuming due to its sequential nature, neither model nor data are typically available at the outset of an experiment. The aim then becomes determining both within a minimal experimental budget. This challenge arises in both manual and high-throughput synthesis contexts. While manual approaches can feasibly explore 1- or 2-dimension (D) parameter spaces (such as binary or ternary phase diagrams), automated synthesis can tackle 2-4 D spaces through batch updates. It's important to note, however, that methods like grid search and data-driven Bayesian Optimization (BO) can be limited in higher dimensions.

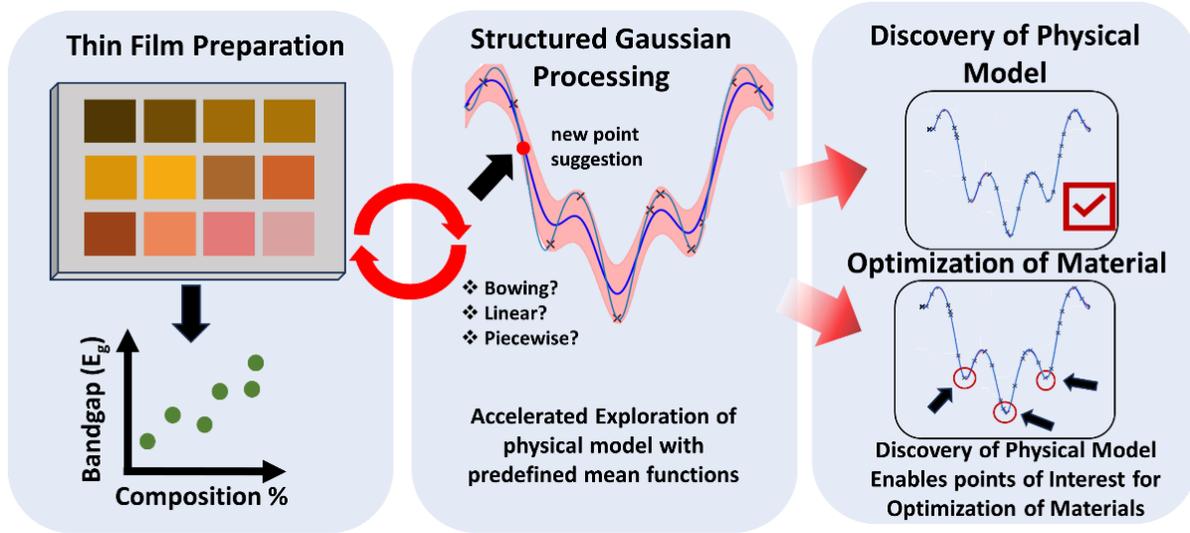

*Figure 1: Workflow for discovery of the physical behavior of bandgap of a mixed perovskite system.*

Recently, we have demonstrated that the introduction of the mean function representing the physical behavior of the system allows accelerating the BO optimization and joint discovery of the physical model and materials optimization during active experiment.[a],[44] Here, we explore the structured Gaussian Process (sGP) approach to discover the physical model of bandgap diagram of $MA_{1-x}GA_xPb(I_{1-x}Br_x)_3$ binary system. The general workflow is shown in the schematic workflow in **Figure 1**. The first step is to prepare a few thin films that can be added into the sGP algorithm. From there the algorithm will suggest a new point to synthesize and measure. This process will repeat until the physical model is discovered and areas of optimization.

**Discussion**

Gaussian Process (GP) is a general class of the non-parametric reconstruction method.[45-47] In the classical GP setting, it is assumed the behavior of the system over a certain parameter space $x$ is given by the ground truth function $f(x)$. This function is unavailable for observers. However, available for observers are the measurements, $y_i$ at the chosen points $x_i$. It is assumed that the relationship between the measurements and the function are given by $y_i = f(x_i)$ + noise, where the noise term is assumed to follow specific (usually Gaussian) distribution. More complex methods allow the noise distribution to be non-Gaussian and change over the parameter space (heteroscedastic noise). The GP aims to reconstruct the function $\hat{f}(x)$, its uncertainty $\sigma(x)$, and noise from the set of observations $(x_i, y_i)$.

The fundamental assumption in GP method is that the values of the function $\hat{f}(x)$ across the parameter space are correlated, and the correlations are described by the kernel function. The kernel function is assumed to have a certain functional form, for example squared exponential or

---
[a] https://github.com/ziatdinovmax/gpax

Matern. For these functions, the kernel is characterized by a parameter vector corresponding to characteristic length scale, and these kernel parameters are learned self-consistently from the data jointly with the noise. Over the last decades, more complex correlation structures represented by possible kernel functions have been extensively explored. However, the characteristic aspect of classical GP methods is that the mean value of the prior function is zero.[48] Classically, this assumption is made based on perceived lack of domain specific prior knowledge of system behavior over the parameter space.

The unique aspect of GP is that they can be used as a foundational framework for active learning methods, e.g. for the development of optimal measurement strategies. In Bayesian Optimization (BO), the predicted function and its uncertainty are combined in the acquisition function balancing the expected gain and potential to discover useful behaviors in weakly explored parts of the parameter space. To accomplish this goal, the human operator suggests the policy balancing the prediction $\hat{f}(x)$ and uncertainty $\sigma(x)$, of the surrogate function. BO is typically implemented as myopic optimization, meaning that the policy is used to plan only the next experimental step, and is implemented as an acquisition function. For example, in greedy approach the next measurement point is chosen as argmax($f(x)$), i.e. the strategy aims to discover the maximum of the function. Alternatively, in pure exploration policy argmax($\sigma(x)$) is chosen as the next measurement point to minimize the uncertainty of the system fastest. Other acquisition functions such as expected improvement (EI) or upper confidence bound (UCB) balance uncertainty and optimization. Note that BO allows for fully probabilistic methods as well, where the search point is drawn from the full posterior distribution (Thompson sampling).

Over the last three years, GP and BO has become the technique of choice for guiding automated experiments in areas as diverse as materials synthesis[49-51], automated microscopy,[52-54] and other instrumental methods.[55-57] However, the BO methods based on pure GP are data driven, meaning that the surrogate model is based purely on data and does not take into account the (partially) known physics of the system. In comparison, classical physical paradigm will rely on constructing the predictive model based on expected physics of the system, starting with the linear approximation, and proceeding to more complex descriptions if evidenced by data.

Recently, we proposed that addition of the probabilistic mean function to the GP can significantly accelerate the discovery.[44, 58] Here, we construct the mean function that represents the possible physics of the process, including the functional form and the priors on parameter distributions. We note that in physical experiments the operator typically has a good idea on the possible prior distributions, either from the previously published work or domain-specific knowledge. The update of the model then updates both the kernel and noise parameters, and the distribution of the model parameters. This approach can be further extended to the scenarios when several models of system behavior, e.g. several structured models and potentially zero mean model, are available, as exemplified in hypothesis learning.[58, 59] The three possible mean functions that are initially defined are linear, quadratic and piecewise which the use of these models will be denoted as using the default sGP functions (d-sGP). Regardless, of the actual physical model d-sGP should be able to discover the ground truth better than the GP algorithm. Furthermore, creating custom mean functions with prior knowledge of physical models can also further accelerate the discovery process. This is where a custom sGP mean function (c-sGP) can be implemented to accelerate discovery faster than d-sGP. Note that in sGP, the physical model is learned during the active experiment along with materials optimization.

With sGP, we can further define several possible tasks for Bayesian Optimization, namely minimization of the model parameter uncertainty, discover the right model, achieve the desired bandgap value, or maximize of minimize bandgap. Here, we first illustrate the discovery of the unknown function using GP based active learning. The underpinning principle of all Bayesian methods is to start with some initial beliefs on the system (priors), acquire data and update beliefs given from the data to get posteriors. The posteriors can be used to estimate the unknown function and its uncertainty. The active learning in GP is realized by creating the acquisition function that balances the estimated function and uncertainty into a single discovery policy (exploration-exploitation tradeoff).

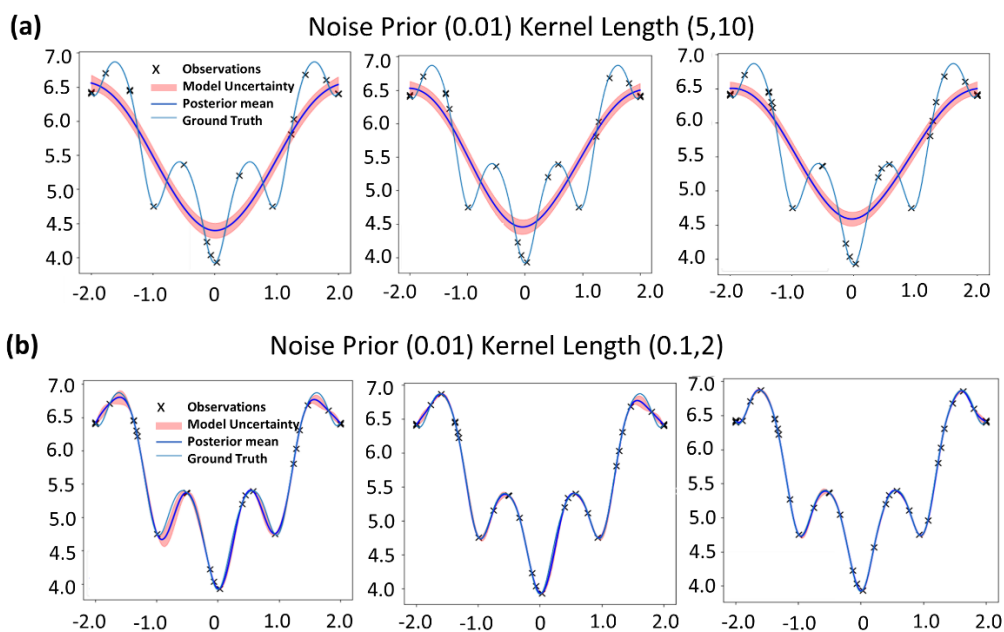

*Figure 2: GP with different noise prior and kernel length (a) Noise prior (0.01) and Kernel length (5,10) (b) noise Prior (0.01) and Kernel length (0.1,2).*

The important aspect of the GP methods is their sensitivity to the initial priors defined before the experiment. While often irrelevant for long experimental sequences when intrinsic properties of the data generation process are discovered, this consideration becomes preponderantly important for short experimental budgets. For a simple GP, the initial parameters for defined functional form of the kernel include the kernel length scale (or lateral scale of correlations in the system, scale (vertical scale), and noise prior. Here, we illustrate the discovery process of the sufficiently complex 1D function using different combinations of these parameters.

The ground truth function is characterized by multiple local minima, accompanied by a noisy component (0.01) (**Figure 2**). The noise prior encompasses the combined effect of our confidence in the precision of our measurements and any uncontrollable factors that might influence the measurements. Similarly, the kernel length scale reflects our belief regarding the potential variation in the properties of the system. If we establish a prior with incorrect or inflexible assumptions, the convergence of the algorithm with the ground truth becomes unattainable. As the algorithm persists in compensating for its inability to accurately describe the ground truth, it eventually loses faith in the provided measurements (**Figure 2a**).

Conversely, if we possess accurate assumptions regarding the noise prior and kernel length, the experimental process can be expedited and simplified. This allows the GP algorithm to rapidly converge towards the ground truth (**Figure 2b**). While GP has the potential to uncover the ground truth given appropriate priors, sGP incorporates a constructed mean function that represents the plausible physics underlying the ground truth. Consequently, sGP facilitates quicker convergence by leveraging this additional information.

In the GP algorithm, the availability of prior data such as prior noise, kernel length, and kernel scale plays a crucial role in reconstructing the function and estimating its uncertainty. However, this process heavily relies on the experimenter's intuition regarding the experiment and its anticipated outcome. The GP algorithm can only discover the true underlying function if the chosen priors enable it to do so. Furthermore, the selection of priors can also influence the number of exploration steps required to converge to the ground truth. To evaluate the performance of the d-sGP mean functions, we examine a toy model labeled as Model 4 (shown in **Figure S2**). This model exhibits an evolution in bandgap, similar to the conditions observed during alloy formation. Specifically, when the variable "x" is small (x < 20% or x > 80%), no alloy is formed. In this model, there exist two regions where alloy formation does not occur, providing us with test cases to assess the convergence capabilities of d-sGP mean functions in capturing the ground truth. Initially, the model is presented with only two data points to make its determination, followed by a total of 30 exploration steps. We anticipate that a more effective model will exhibit a faster decrease in the overall uncertainty within the system.

d-sGP require the incorporation of physics or domain knowledge to effectively model a system. However, it is important to exercise caution when employing priors devoid of physical foundations, as blind adherence to such priors can lead to prolonged experimental timelines. In our investigation (**Figure S2a**), we utilized a linear mean function ($y = a \times x + b$) to explore a toy Model 4. The priors assigned to the linear equation introduced a large variance (0,1000) for the slope coefficient (a), indicating a normal distribution centered around 0 with a standard deviation of 1000. Similarly, the prior for the intercept (b) was defined as a normal distribution (1000,1000), favoring larger intercept values.

Upon commencing a 30-step exploration, we observed that the algorithm initially relied on the given priors and failed to discover areas where alloy formation did not occur before step 20. However, after surpassing this threshold, the algorithm progressively disregarded the priors and embarked on a more comprehensive exploration of the system. It is worth noting that priors should not be perceived as optimization objectives, but rather as intuitive guesses informed by our understanding of the data.

In **Figure S2b**, we employed a log-normal distribution with mean 0 and variance 1 for the slope, allowing for the discovery of positive slopes and capturing the concentration range more effectively. The intercept modeled using a Normal distribution (0,2), exhibited symmetry around its mean. Notably, by step 10, the algorithm began uncovering the change in alloy formation around 80%, while by step 20, it successfully identified the alteration occurring below 20%. This highlights how the choice of priors can significantly impact the duration of an experiment, underscoring their influence on the efficiency of the discovery process. The incorporation of appropriate mean function priors in d-sGPs facilitates the capture of expected and unexpected system behaviors. However, the selection of priors should be guided by our knowledge and intuition rather than viewing them as mere optimization targets. By making informed choices, we can expedite the discovery process and reduce experimental timelines, ultimately enhancing our understanding of the system at hand.

Here, we first explore this approach for the simplified model with the known ground truth. The detailed description is provided in the Google Colab tutorial (found in supplementary under Data Availability) and can be reproduced in full or adapted to different models. As an illustration, we have created a toy model that is piecewise with a quadratic interval and a linear interval (**Figure S1**). Accessible to the experimentalist is the choice of the mean function of the BO. We consider the case where $m = 0$, i.e. classical GP, the cases where the mean function is linear, piecewise, and quadratic and a custom mean function (piecewise quadratic) is defined given prior knowledge of our toy model. By definition of the probabilistic models, the choice of function is complemented by the definition of the prior distribution of the parameters of the function, kernel length and scale, and noise. Based upon experimentation, we observed optimal behavior in the scenarios when the distributions are sufficiently broad to accommodate realistic behaviors but are still sufficiently narrow to avoid the lengthy initial convergence before correcting values are identified. Practically, the noise level prior is our guess of how noisy our data is or how much we trust our measurements. Here the noise of our toy model is 0.01 and our noise prior is 0.1 meaning that our noise is large, and we do not trust our data. Shown in **Figure 3** is the evolution of the vanilla GP for our toy model, sGP with a linear mean function and custom mean function for 30 exploration steps.

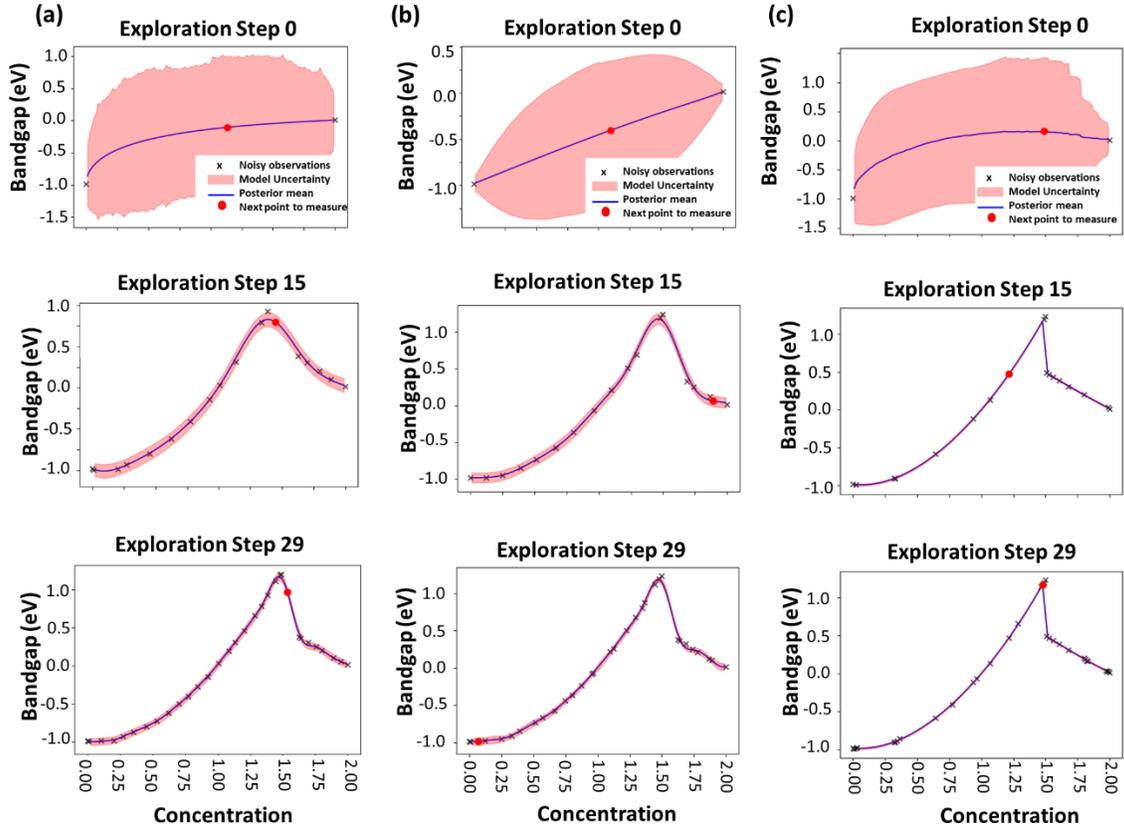

*Figure 3: (a) GP (b) d-sGP with linear mean function c) c-sGP with a piecewise quadratic mean function for piecewise quadratic toy model for 30 exploration steps*

The GP model (**Figure 3a**) commences with a high level of uncertainty, and with each exploration step, the uncertainty gradually decreases, albeit at a slower pace compared to the d-sGP and c-sGP models. Throughout the exploration process, the GP model has a fairly high uncertainty value around 25 after between steps 5 and 25 shown in **Figure 4**. On the other hand, the d-sGP model depicted in **Figure 3b** displays a rapid reduction in uncertainty by step 15 in comparison to the GP model. By the end of the 30 exploration steps, the d-sGP model has not yet converged to the ground truth but has provided valuable insights regarding the known ground truth. Given our knowledge of the ground truth, the c-sGP employed in **Figure 3c** was able to discover the ground truth before exploration step 15 and has uncertainty values around 0 by the end of the 30 explorations (**Figure 4**). Due to our knowledge of the ground truth, we are also able to look at the deviation between the predicted posterior mean and ground truth, mean squared error (MSE), which allows us to determine how our prediction is off from the ground truth. All the models in **Figure 4** have fairly low MSE meaning they have close correlation between the predicted posterior mean and ground truth, but we can see that the quadratic and piecewise mean functions increase in correlation between the predicted posterior mean and ground truth over the exploration steps. Yet, the best performing mode was c-sGP having an MSE of 0 from step 7. In summary, the comparison between the GP, d-sGP and c-sGP mean functions reveals that the c-sGP model demonstrates a more efficient uncertainty reduction process within the given exploration steps.

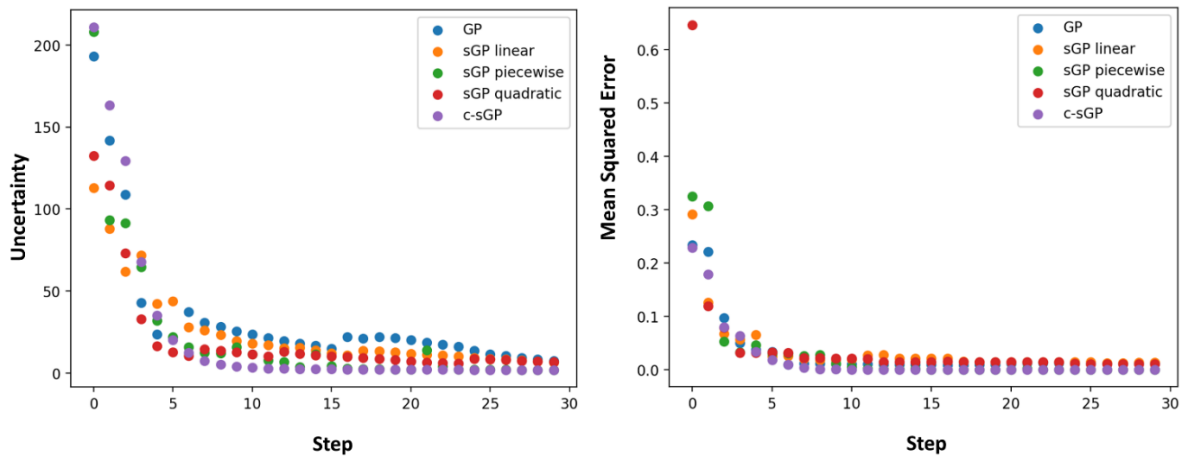

*Figure 4: uncertainty and MSE of piecewise quadratic toy model with comparisons of GP and d-sGP and c-sGP with mean functions linear, quadratic, piecewise and custom piecewise quadratic.*

Henceforth, we employed the c-sGP methodology to analyze a 1D binary dataset comprising an addition of GAPbBr$_3$ to MAPbI$_3$. Our objective was to examine the efficiency of the c-sGP mean function in accurately determining the true nature of this system. Since the ground truth is inaccessible when working with real-world data, we resorted to experimental investigations in order to elucidate the relationship between the addition of GAPbBr$_3$ and MAPbI$_3$. For the custom mean function, bandgap toy models (**Figure S2**) which are based on known bandgap model formations are combined into one custom function. Since the ground truth of our experimental data is unknown and not all defined models may be applicable to the experiment each bandgap model is given a categorial variable that will represent the 'weight' of each model. The weights help decide the influence of each model based on the data. The model with a higher weight will have a higher contribution to the discovery in comparison to the model with a lower weight. The final output will be the average of all the weights. The weights are determined using the Dirichlet distribution which models the probability of k events that are mutually exclusive and collectively exhaustive. This makes it perfect for model mixing occurring in this c-sGP algorithm. To ascertain the mean function exhibiting the least amount of model uncertainty, we conducted a comparative evaluation involving two initial compositions and subsequently expanded the study to encompass a total of thirty compositions (**Figure 5a-c**). Similarly to GP, d-sGP and c-sGP priors also hold importance in effectively capturing the underlying patterns in the data. In order to adequately account for the inherent noise in our data, we chose a noise level of 0.01, striking a balance between a low and high noise assumption.

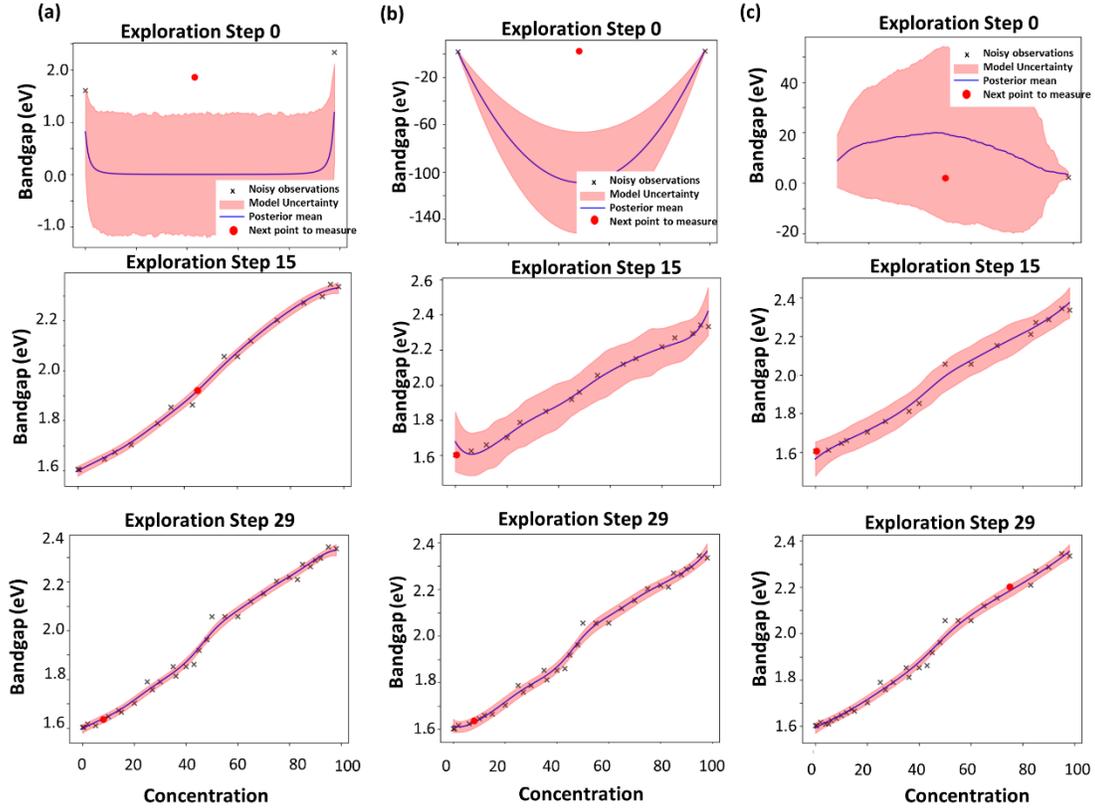

*Figure 5: 30 Step exploration of (a) Custom (b) Quadratic (c) Piecewise mean functions*

In the analysis pertaining to the custom function depicted in **Figure 5a**, it is observed that the uncertainty undergoes a rapid reduction within the initial 15 exploratory iterations. During this phase, the uncertainty commences at an amplitude of 270, subsequently declining to approximately 3 by the 15th iteration, as evidenced in **Figure S4**. Contrarily, both the quadratic and piecewise functions, depicted in **Figures 5b-c**, continue to manifest pronounced uncertainty magnitudes by the same iteration count. Specifically, the quadratic function initiates with an uncertainty of 5702, reducing to 23 by the aforementioned step, while the piecewise function, beginning at 4720, diminishes to 14. An examination of the scatter plot presented in **Figure S4** indicates superior performance of the c-sGP and Linear d-sGP functions. However, it is imperative to note that the uncertainty inherent to the models should not be the sole criterion for model efficacy; the observations derived from exploratory iterations are equally important. The experimental model exhibits non-linear characteristics, marked by sporadic escalations in bandgaps within specific $GAPbBr_3$ inclusion ranges, such as sub-20% and interstitially between 50-60%. The d-sGP and c-sGP models, interestingly, do not capture all data points pertaining to the measured bandgaps. This deviation may be attributed to the algorithm's tendency to classify certain variations as noise, particularly when data points diverge from the predominant trend exhibited by other points within the dataset.

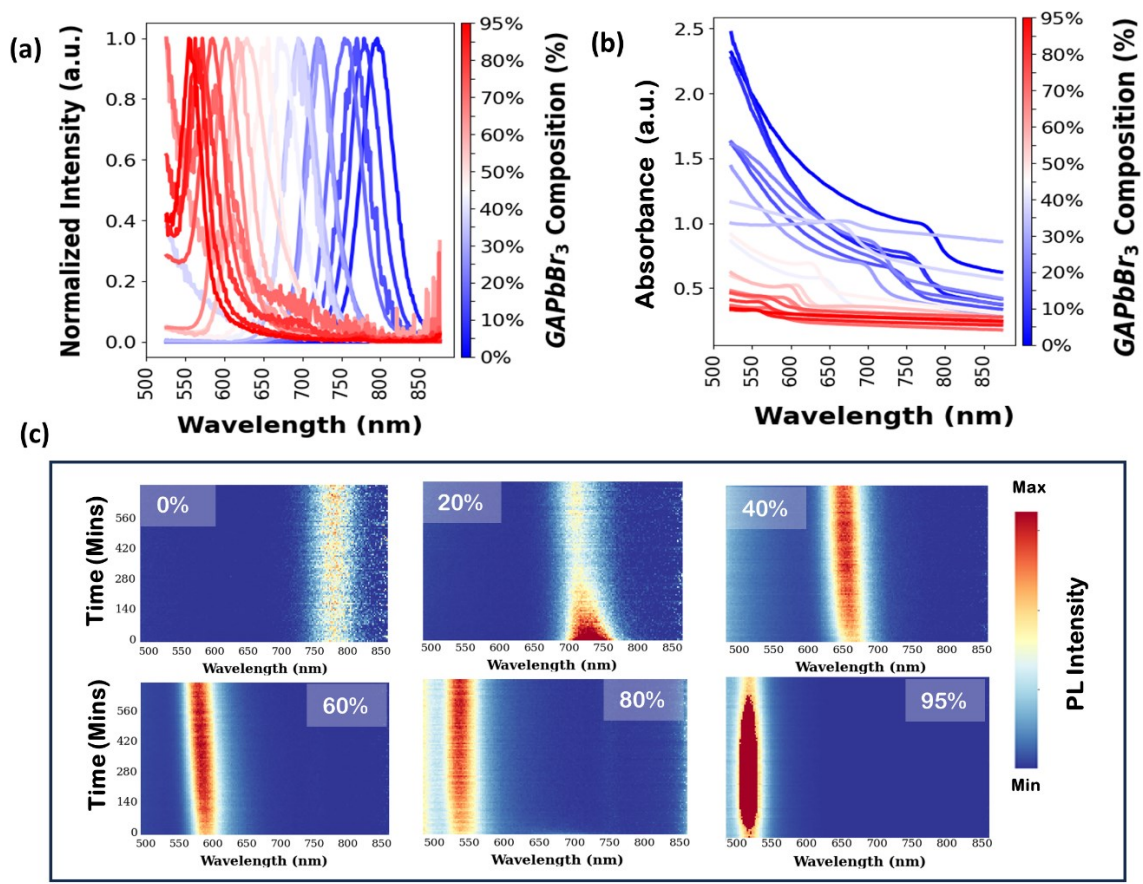

*Figure 6: The experimental data for 20 mixed MAPbI$_3$ and GAPbBr$_3$ with various concentration. (a) Normalize PL (b) absorbance spectra (c) Time dependent photoluminescence spectra heatmaps with various additions of GAPbBr$_3$ to MAPbI$_3$*

The influence exerted by the addition of GAPbBr$_3$ can be readily seen through the discernible augmentation of both the photoluminescence (PL) and absorbance, as illustrated in **Figure 6b-c.** Notably, there is an increase in intensity with low addition of GAPbBr$_3$ (2 to 6%) shown in **Figure S5b** where the PL peak doesn't have an obvious shift. This may be attributed to the passivation action of the GAPbBr$_3$ towards the defect at the MAPbI$_3$ surface,here the ions are not permeated towards the bulk perovskite lattice.[60] However, a change in this observed trend becomes apparent upon further increments in the concentration of GAPbBr$_3$, wherein the PL experiences a blue-shift and a concurrent decrease in intensity up to 80% (**Figure 6a and S5a**). The origin of this peak shift may be due to bandgap variation due to halide substitution. MAPbI$_3$ has a relatively lower bandgap, which has a PL peak around the 750 nm region. On the other hand, GAPbBr$_3$, being bromide-based, has a higher bandgap which translates to a PL peak in the blue-region. This leads to a gradual blue shift of the PL peak.[43, 61] This decrease in intensity can possibly be attributed to various reasons such as the crystal structure evolution where the mixed-halide system can have halide rich regions that can create non-radiative recombination pathways.[62, 63] Also as the bromide content increases, the bandgap is widening which can impact the PL intensity.[24, 61] Moreover, a distinct alteration in the absorption spectrum is observed with the introduction of GAPbBr$_3$. Specifically, the characteristic absorbance peak, localized around 800 nm, vanishes as the concentration of GAPbBr$_3$ increases, being replaced by new peaks that manifest in the range of 640 nm to 550 nm when approximately 50% of GAPbBr$_3$ is introduced (**Figure 6b**). To validate the absence of phase segregation, the PL was measured over a continuous duration of 12 hours under ambient conditions, as depicted in **Figure 6c**. The PL changes over time was shown for 0, 20, 40, 60, 80 and 95% GAPbBr$_3$. We can see over time the intensity of pure MAPbI$_3$ remains the same without any PL peak shifts. As we add a higher amount of GAPbBr$_3$ there is evidence of blue-shifting over time for 20% addition and at 40% over the 12 hours there are no apparent PL peak shifts and the intensity remains high for the total time. We see a systematic shift in PL peak as we continue to add GAPbBr$_3$ with an increase in PL intensity. The use of GA ions have reported to passivate defects in the perovskite systems, which can reduce non-radiative recombination's.[41, 43, 64] At the higher concentrations of GAPbBr$_3$ there may be a possibility of passivation of defects without causing phase segregation leading to an increase in PL. The ensuing thin film observations, as depicted in **Figure S5c**, show that pure MAPbI$_3$ films have a medium brown color. When low amount of GAPbBr$_3$ (below 12%) is added the films turn into a deepened dark brown color, then a lighter brown color at higher quantities of GAPbBr$_3$. Addition of GAPbBr$_3$ has also been shown to increase the intensity of the PL in tandem with a small addition of GAPbBr$_3$ (as observed in **Figure S5b**). It is evident that concentrations below the 20% threshold exhibit minimal alterations in the quantified bandgap, no halide segregation or phase transition and seemly increases the PL intensity. At 20% the bandgap has increased from 1.604 eV to 1.703 eV (**Figure s6**) and remains around 1.7 eV up to 40%. After this addition percentage the bandgap jumps up to 2.0 eV and remains around there when increasing GAPbBr$_3$ component.

In comparison, phase segregation in mixed halide perovskites, such as in MAPbI$_3$ and MAPbBr$_3$, becomes particularly pronounced when exposed to ambient conditions.[62, 65, 66] Under environmental influences such as light[9, 63, 65] and moisture,[66, 67] the mixed halide perovskite tends to separate into its individual components. Phase segregation can induce localized regions of iodide-rich or bromide-rich domains. This process will reduce the overall homogeneity of the mixed material and can compromise its optoelectronic properties. Incorporating GA into the A-site cation in MA-based hybrid perovskite has shown enhanced stability and even better air stability than MAPbI$_3$.[68]

The blending of mixed halide perovskite materials tends to result in bandgap behavior that is characterized by different models based on the specific components being mixed. For example, mixing halides such as iodide, bromide and chloride tend to result in a linear model.[24, 69] Whereas, mixing the divalent cation such as lead and tin exhibits a bowing model.[70-72] In this case, we see some apparent jumps or anomalies in bandgap, especially above 40% addition of GAPbBr$_3$ which can possibly be influenced by certain factors. Such as, GA being large enough that it can't exactly occupy the A-site cation between the [PbI$_6$]$^{4-}$ arrays, causing induced local lattice distortions.[41, 73] Also, pure GAPbBr$_3$ doesn't exist in a 3D phase which the introduction of GAPbBr$_3$ to MAPbI$_3$ may induce some structural changes or phase transitions leading to this variation in bandgap.[74]

Overall, utilizing sGP allow us to not only identify some underlying physical model, but also pinpoint areas for optimization. The ability to discern intricate relationships and its capability to parse complex material datasets shows its power in material discovery and optimization.

In this study, we have successfully applied the aforementioned workflow to the sequential process of thin film preparation. This methodology can easily be extended to accommodate automated synthesis workflows such as microfluidics, pipetting robots,[75-79] and fully autonomous labs. Incorporation of higher dimensional spaces and implementing batch updates allows this workflow to be adaptable to a broader range of experimental scenarios. To expedite BO and facilitate the simultaneous discovery of the physical model, we have developed a novel algorithm. This algorithm offers a significant acceleration in the optimization process, allowing for more efficient exploration and exploitation of the parameter space. The mean function, which is a central component of our algorithm, embodies the physical behaviors and characteristics of the system by capturing the essential features and potential mechanisms that drive discovery and material optimization during the active experiment. This joint exploration of the parameter space and active discovery of the physical model contributes to the advancement of scientific understanding and optimization of material properties. Overall, our approach demonstrates the effectiveness of incorporating the mean function, which encapsulates the physical behaviors of the system. This integration enhances the efficiency of the optimization process, facilitating more effective and targeted experiments for the discovery and optimization of materials.

Acknowledgements:

S.S., E.F., J.Y. and M.A. acknowledge support from the National Science Foundation (NSF), award number 2043205. The authors acknowledge support from the Center for Nanophase Materials Sciences (CNMS) user facility, project CNMS2023-A-01927, which is a US Department of Energy, Office of Science User Facility at Oak Ridge National Laboratory.

# References

(1) Kojima, A.; Teshima, K.; Shirai, Y.; Miyasaka, T. Organometal Halide Perovskites as Visible-Light Sensitizers for Photovoltaic Cells. *Journal of the American Chemical Society* **2009**, *131* (17), 6050-6051. DOI: 10.1021/ja809598r.
(2) Park, J.; Kim, J.; Yun, H.-S.; Paik, M. J.; Noh, E.; Mun, H. J.; Kim, M. G.; Shin, T. J.; Seok, S. I. Controlled growth of perovskite layers with volatile alkylammonium chlorides. *Nature* **2023**, *616* (7958), 724-730. DOI: 10.1038/s41586-023-05825-y.
(3) Mariotti, S.; Köhnen, E.; Scheler, F.; Sveinbjörnsson, K.; Zimmermann, L.; Piot, M.; Yang, F.; Li, B.; Warby, J.; Musiienko, A.; et al. Interface engineering for high-performance, triple-halide perovskite–silicon tandem solar cells. *Science* **2023**, *381* (6653), 63-69. DOI: doi:10.1126/science.adf5872.
(4) Kung, P. K.; Li, M. H.; Lin, P. Y.; Chiang, Y. H.; Chan, C. R.; Guo, T. F.; Chen, P. A review of inorganic hole transport materials for perovskite solar cells. *Advanced Materials Interfaces* **2018**, *5* (22), 1800882.
(5) Kubicki, D. J.; Prochowicz, D.; Hofstetter, A.; Saski, M.; Yadav, P.; Bi, D.; Pellet, N.; Lewinski, J.; Zakeeruddin, S. M.; Gratzel, M.; et al. Formation of Stable Mixed Guanidinium-Methylammonium Phases with Exceptionally Long Carrier Lifetimes for High-Efficiency Lead Iodide-Based Perovskite Photovoltaics. *J Am Chem Soc* **2018**, *140* (9), 3345-3351. DOI: 10.1021/jacs.7b12860 From NLM PubMed-not-MEDLINE.
(6) He, J.; Fang, W.-H.; Long, R.; Prezhdo, O. V. Increased Lattice Stiffness Suppresses Nonradiative Charge Recombination in MAPbI3 Doped with Larger Cations: Time-Domain Ab Initio Analysis. *ACS Energy Letters* **2018**, *3* (9), 2070-2076. DOI: 10.1021/acsenergylett.8b01191.
(7) Pham, N. D.; Zhang, C.; Tiong, V. T.; Zhang, S.; Will, G.; Bou, A.; Bisquert, J.; Shaw, P. E.; Du, A.; Wilson, G. J.; et al. Tailoring Crystal Structure of FA0.83Cs0.17PbI3 Perovskite Through Guanidinium Doping for Enhanced Performance and Tunable Hysteresis of Planar Perovskite Solar Cells. *Advanced Functional Materials* **2019**, *29* (1), 1806479. DOI: https://doi.org/10.1002/adfm.201806479.
(8) Yang, W.; Long, H.; Sha, X.; Sun, J.; Zhao, Y.; Guo, C.; Peng, X.; Shou, C.; Yang, X.; Sheng, J.; et al. Unlocking Voltage Potentials of Mixed-Halide Perovskite Solar Cells via Phase Segregation Suppression. *Advanced Functional Materials* **2022**, *32* (12), 2110698. DOI: https://doi.org/10.1002/adfm.202110698.
(9) Datta, K.; van Gorkom, B. T.; Chen, Z.; Dyson, M. J.; van der Pol, T. P. A.; Meskers, S. C. J.; Tao, S.; Bobbert, P. A.; Wienk, M. M.; Janssen, R. A. J. Effect of Light-Induced Halide Segregation on the Performance of Mixed-Halide Perovskite Solar Cells. *ACS Applied Energy Materials* **2021**, *4* (7), 6650-6658. DOI: 10.1021/acsaem.1c00707.
(10) Hoke, E. T.; Slotcavage, D. J.; Dohner, E. R.; Bowring, A. R.; Karunadasa, H. I.; McGehee, M. D. Reversible photo-induced trap formation in mixed-halide hybrid perovskites for photovoltaics. *Chemical Science* **2015**, *6* (1), 613-617, 10.1039/C4SC03141E. DOI: 10.1039/C4SC03141E.
(11) Liu, Y.; Wang, M.; Ievlev, A. V.; Ahmadi, A.; Keum, J. K.; Ahmadi, M.; Hu, B.; Ovchinnikova, O. S. Photoinduced iodide repulsion and halides-demixing in layered perovskites. *Materials Today Nano* **2022**, *18*, 100197. DOI: https://doi.org/10.1016/j.mtnano.2022.100197.
(12) Adjokatse, S.; Fang, H.-H.; Loi, M. A. Broadly tunable metal halide perovskites for solid-state light-emission applications. *Materials Today* **2017**, *20* (8), 413-424. DOI: 10.1016/j.mattod.2017.03.021.
(13) Straus, D. B.; Cava, R. J. Tuning the Band Gap in the Halide Perovskite CsPbBr(3) through Sr Substitution. *ACS Appl Mater Interfaces* **2022**, *14* (30), 34884-34890. DOI: 10.1021/acsami.2c09275 From NLM PubMed-not-MEDLINE.
(14) Hassan, Y.; Park, J. H.; Crawford, M. L.; Sadhanala, A.; Lee, J.; Sadighian, J. C.; Mosconi, E.; Shivanna, R.; Radicchi, E.; Jeong, M.; et al. Ligand-engineered bandgap stability in mixed-halide perovskite LEDs. *Nature* **2021**, *591* (7848), 72-77. DOI: 10.1038/s41586-021-03217-8.
(15) Ahn, C. W.; Jo, J. H.; Choi, J. S.; Hwang, Y. H.; Kim, I. W.; Kim, T. H. Heteroanionic Lead-Free Double-Perovskite Halides for Bandgap Engineering. *Advanced Engineering Materials* **2023**, *25* (1), 2201119. DOI: https://doi.org/10.1002/adem.202201119.

(16) Wen, J.; Zhao, Y.; Liu, Z.; Gao, H.; Lin, R.; Wan, S.; Ji, C.; Xiao, K.; Gao, Y.; Tian, Y.; et al. Steric Engineering Enables Efficient and Photostable Wide-Bandgap Perovskites for All-Perovskite Tandem Solar Cells. *Advanced Materials* **2022**, *34* (26), 2110356. DOI: https://doi.org/10.1002/adma.202110356.
(17) Byranvand, M. M.; Otero-Martínez, C.; Ye, J.; Zuo, W.; Manna, L.; Saliba, M.; Hoye, R. L. Z.; Polavarapu, L. Recent Progress in Mixed A-Site Cation Halide Perovskite Thin-Films and Nanocrystals for Solar Cells and Light-Emitting Diodes. *Advanced Optical Materials* **2022**, *10* (14), 2200423. DOI: https://doi.org/10.1002/adom.202200423.
(18) Kim, G.; Min, H.; Lee, K. S.; Lee, D. Y.; Yoon, S. M.; Seok, S. I. Impact of strain relaxation on performance of α-formamidinium lead iodide perovskite solar cells. *Science* **2020**, *370* (6512), 108-112. DOI: doi:10.1126/science.abc4417.
(19) Duong, T.; Nguyen, T.; Huang, K.; Pham, H.; Adhikari, S. G.; Khan, M. R.; Duan, L.; Liang, W.; Fong, K. C.; Shen, H.; et al. Bulk Incorporation with 4-Methylphenethylammonium Chloride for Efficient and Stable Methylammonium-Free Perovskite and Perovskite-Silicon Tandem Solar Cells. *Advanced Energy Materials* **2023**, *13* (9), 2203607. DOI: https://doi.org/10.1002/aenm.202203607.
(20) Wang, H.; Liu, H.; Dong, Z.; Li, W.; Zhu, L.; Chen, H. Composition manipulation boosts the efficiency of carbon-based CsPbI3 perovskite solar cells to beyond 14%. *Nano Energy* **2021**, *84*, 105881. DOI: https://doi.org/10.1016/j.nanoen.2021.105881.
(21) Lee, J.-W.; Tan, S.; Seok, S. I.; Yang, Y.; Park, N.-G. Rethinking the A cation in halide perovskites. *Science* **2022**, *375* (6583), eabj1186. DOI: doi:10.1126/science.abj1186.
(22) Amat, A.; Mosconi, E.; Ronca, E.; Quarti, C.; Umari, P.; Nazeeruddin, M. K.; Grätzel, M.; De Angelis, F. Cation-Induced Band-Gap Tuning in Organohalide Perovskites: Interplay of Spin–Orbit Coupling and Octahedra Tilting. *Nano Letters* **2014**, *14* (6), 3608-3616. DOI: 10.1021/nl5012992.
(23) Higgins, K.; Valleti, S. M.; Ziatdinov, M.; Kalinin, S. V.; Ahmadi, M. Chemical Robotics Enabled Exploration of Stability in Multicomponent Lead Halide Perovskites via Machine Learning. *ACS Energy Letters* **2020**, *5* (11), 3426-3436. DOI: 10.1021/acsenergylett.0c01749.
(24) Noh, J. H.; Im, S. H.; Heo, J. H.; Mandal, T. N.; Seok, S. I. Chemical Management for Colorful, Efficient, and Stable Inorganic–Organic Hybrid Nanostructured Solar Cells. *Nano Letters* **2013**, *13* (4), 1764-1769. DOI: 10.1021/nl400349b.
(25) Sutter-Fella, C. M.; Li, Y.; Amani, M.; Ager, J. W., III; Toma, F. M.; Yablonovitch, E.; Sharp, I. D.; Javey, A. High Photoluminescence Quantum Yield in Band Gap Tunable Bromide Containing Mixed Halide Perovskites. *Nano Letters* **2016**, *16* (1), 800-806. DOI: 10.1021/acs.nanolett.5b04884.
(26) Green, M.; Dunlop, E.; Hohl-Ebinger, J.; Yoshita, M.; Kopidakis, N.; Hao, X. Solar cell efficiency tables (version 57). *Progress in Photovoltaics: Research and Applications* **2020**, *29* (1), 3-15. DOI: 10.1002/pip.3371.
(27) Suchan, K.; Just, J.; Beblo, P.; Rehermann, C.; Merdasa, A.; Mainz, R.; Scheblykin, I. G.; Unger, E. Multi-Stage Phase-Segregation of Mixed Halide Perovskites under Illumination: A Quantitative Comparison of Experimental Observations and Thermodynamic Models. *Advanced Functional Materials* **2023**, *33* (3), 2206047. DOI: https://doi.org/10.1002/adfm.202206047.
(28) Kahmann, S.; Chen, Z.; Hordiichuk, O.; Nazarenko, O.; Shao, S.; Kovalenko, M. V.; Blake, G. R.; Tao, S.; Loi, M. A. Compositional Variation in FAPb(1-x)Sn(x)I(3) and Its Impact on the Electronic Structure: A Combined Density Functional Theory and Experimental Study. *ACS Appl Mater Interfaces* **2022**, *14* (30), 34253-34261. DOI: 10.1021/acsami.2c00889 From NLM.
(29) Goyal, A.; McKechnie, S.; Pashov, D.; Tumas, W.; van Schilfgaarde, M.; Stevanović, V. Origin of Pronounced Nonlinear Band Gap Behavior in Lead–Tin Hybrid Perovskite Alloys. *Chemistry of Materials* **2018**, *30* (11), 3920-3928. DOI: 10.1021/acs.chemmater.8b01695.
(30) Ndione, P. F.; Li, Z.; Zhu, K. Effects of alloying on the optical properties of organic–inorganic lead halide perovskite thin films. *Journal of Materials Chemistry C* **2016**, *4* (33), 7775-7782, 10.1039/C6TC02135B. DOI: 10.1039/C6TC02135B.

(31) Tao, S.; Schmidt, I.; Brocks, G.; Jiang, J.; Tranca, I.; Meerholz, K.; Olthof, S. Absolute energy level positions in tin- and lead-based halide perovskites. *Nature Communications* **2019**, *10* (1), 2560. DOI: 10.1038/s41467-019-10468-7.
(32) Olthof, S. Research Update: The electronic structure of hybrid perovskite layers and their energetic alignment in devices. *APL Materials* **2016**, *4* (9). DOI: 10.1063/1.4960112 (acccessed 9/12/2023).
(33) Emara, J.; Schnier, T.; Pourdavoud, N.; Riedl, T.; Meerholz, K.; Olthof, S. Impact of Film Stoichiometry on the Ionization Energy and Electronic Structure of CH3NH3PbI3 Perovskites. *Advanced Materials* **2016**, *28* (3), 553-559. DOI: https://doi.org/10.1002/adma.201503406.
(34) Yin, Y.; Wang, Y.; Sun, Q.; Yang, Y.; Wang, Y.; Yang, Z.; Yin, W. J. Unique Photoelectric Properties and Defect Tolerance of Lead-Free Perovskite Cs(3)Cu(2)I(5) with Highly Efficient Blue Emission. *J Phys Chem Lett* **2022**, *13* (18), 4177-4183. DOI: 10.1021/acs.jpclett.2c00888 From NLM PubMed-not-MEDLINE.
(35) Subedi, B.; Li, C.; Chen, C.; Liu, D.; Junda, M. M.; Song, Z.; Yan, Y.; Podraza, N. J. Urbach Energy and Open-Circuit Voltage Deficit for Mixed Anion-Cation Perovskite Solar Cells. *ACS Appl Mater Interfaces* **2022**, *14* (6), 7796-7804. DOI: 10.1021/acsami.1c19122 From NLM PubMed-not-MEDLINE.
(36) Basumatary, P.; Kumari, J.; Agarwal, P. Probing the defects states in MAPbI3 perovskite thin films through photoluminescence and photoluminescence excitation spectroscopy studies. *Optik* **2022**, *266*. DOI: 10.1016/j.ijleo.2022.169586.
(37) Liu, Y.; Ievlev, A. V.; Borodinov, N.; Lorenz, M.; Xiao, K.; Ahmadi, M.; Hu, B.; Kalinin, S. V.; Ovchinnikova, O. S. Direct Observation of Photoinduced Ion Migration in Lead Halide Perovskites. *Advanced Functional Materials* **2020**, *31* (8). DOI: 10.1002/adfm.202008777.
(38) Liu, Y.; Borodinov, N.; Lorenz, M.; Ahmadi, M.; Kalinin, S. V.; Ievlev, A. V.; Ovchinnikova, O. S. Hysteretic Ion Migration and Remanent Field in Metal Halide Perovskites. *Adv Sci (Weinh)* **2020**, *7* (19), 2001176. DOI: 10.1002/advs.202001176 From NLM PubMed-not-MEDLINE.
(39) Mahapatra, A.; Runjhun, R.; Nawrocki, J.; Lewiński, J.; Kalam, A.; Kumar, P.; Trivedi, S.; Tavakoli, M. M.; Prochowicz, D.; Yadav, P. Elucidation of the role of guanidinium incorporation in single-crystalline MAPbI3 perovskite on ion migration and activation energy. *Physical Chemistry Chemical Physics* **2020**, *22* (20), 11467-11473, 10.1039/D0CP01119C. DOI: 10.1039/D0CP01119C.
(40) Kubicki, D. J.; Prochowicz, D.; Hofstetter, A.; Saski, M.; Yadav, P.; Bi, D.; Pellet, N.; Lewiński, J.; Zakeeruddin, S. M.; Grätzel, M.; et al. Formation of Stable Mixed Guanidinium–Methylammonium Phases with Exceptionally Long Carrier Lifetimes for High-Efficiency Lead Iodide-Based Perovskite Photovoltaics. *Journal of the American Chemical Society* **2018**, *140* (9), 3345-3351. DOI: 10.1021/jacs.7b12860.
(41) Jodlowski, A. D.; Roldán-Carmona, C.; Grancini, G.; Salado, M.; Ralaiarisoa, M.; Ahmad, S.; Koch, N.; Camacho, L.; de Miguel, G.; Nazeeruddin, M. K. Large guanidinium cation mixed with methylammonium in lead iodide perovskites for 19% efficient solar cells. *Nature Energy* **2017**, *2* (12), 972-979. DOI: 10.1038/s41560-017-0054-3.
(42) Boix, P. P.; Agarwala, S.; Koh, T. M.; Mathews, N.; Mhaisalkar, S. G. Perovskite Solar Cells: Beyond Methylammonium Lead Iodide. *J Phys Chem Lett* **2015**, *6* (5), 898-907. DOI: 10.1021/jz502547f From NLM PubMed-not-MEDLINE.
(43) Balaji Gandhi, M.; Valluvar Oli, A.; Nicholson, S.; Adelt, M.; Martin, R.; Chen, Y.; Babu Sridharan, M.; Ivaturi, A. Investigation on guanidinium bromide incorporation in methylammonium lead iodide for enhanced efficiency and stability of perovskite solar cells. *Solar Energy* **2023**, *253*, 1-8. DOI: https://doi.org/10.1016/j.solener.2023.01.026.
(44) Ziatdinov, M. A.; Ghosh, A.; Kalinin, S. V. Physics makes the difference: Bayesian optimization and active learning via augmented Gaussian process. *Machine Learning: Science and Technology* **2022**, *3* (1), 015003. DOI: 10.1088/2632-2153/ac4baa.
(45) Rasmussen, C. E.; Williams, C. K. I. *Gaussian Processes for Machine Learning (Adaptive Computation and Machine Learning)*; The MIT Press, 2005.
(46) Lambert, B. *A Student's Guide to Bayesian Statistics*; SAGE Publications Ltd; 1 edition, 2018.

(47) Martin, O. *Bayesian Analysis with Python: Introduction to statistical modeling and probabilistic programming using PyMC3 and ArviZ, 2nd Edition*; Packt Publishing, 2018.
(48) Garnett, R. *Bayesian Optimization, https://bayesoptbook.com/*; Cambridge University Press, 2022.
(49) Noack, M. M.; Doerk, G. S.; Li, R.; Streit, J. K.; Vaia, R. A.; Yager, K. G.; Fukuto, M. Autonomous materials discovery driven by Gaussian process regression with inhomogeneous measurement noise and anisotropic kernels. *Scientific Reports* **2020**, *10* (1), 17663. DOI: 10.1038/s41598-020-74394-1.
(50) Ahmadi, M.; Ziatdinov, M.; Zhou, Y.; Lass, E. A.; Kalinin, S. V. Machine learning for high-throughput experimental exploration of metal halide perovskites. *Joule* **2021**, *5* (11), 2797-2822. DOI: https://doi.org/10.1016/j.joule.2021.10.001.
(51) Ament, S.; Amsler, M.; Sutherland, D. R.; Chang, M.-C.; Guevarra, D.; Connolly, A. B.; Gregoire, J. M.; Thompson, M. O.; Gomes, C. P.; van Dover, R. B. Autonomous materials synthesis via hierarchical active learning of nonequilibrium phase diagrams. *Science Advances* **2021**, *7* (51), eabg4930. DOI: doi:10.1126/sciadv.abg4930.
(52) Ziatdinov, M.; Liu, Y.; Kelley, K.; Vasudevan, R.; Kalinin, S. V. Bayesian Active Learning for Scanning Probe Microscopy: From Gaussian Processes to Hypothesis Learning. *ACS Nano* **2022**, *16* (9), 13492-13512. DOI: 10.1021/acsnano.2c05303.
(53) Liu, Y.; Yang, J.; Vasudevan, R. K.; Kelley, K. P.; Ziatdinov, M.; Kalinin, S. V.; Ahmadi, M. Exploring the Relationship of Microstructure and Conductivity in Metal Halide Perovskites via Active Learning-Driven Automated Scanning Probe Microscopy. *J Phys Chem Lett* **2023**, *14* (13), 3352-3359. DOI: 10.1021/acs.jpclett.3c00223  From NLM.
(54) Liu, Y.; Kelley, K. P.; Vasudevan, R. K.; Funakubo, H.; Ziatdinov, M. A.; Kalinin, S. V. Experimental discovery of structure–property relationships in ferroelectric materials via active learning. *Nature Machine Intelligence* **2022**, *4* (4), 341-350. DOI: 10.1038/s42256-022-00460-0.
(55) Boelrijk, J.; Pirok, B.; Ensing, B.; Forré, P. Bayesian optimization of comprehensive two-dimensional liquid chromatography separations. *Journal of Chromatography A* **2021**, *1659*, 462628. DOI: https://doi.org/10.1016/j.chroma.2021.462628.
(56) Boelrijk, J.; Ensing, B.; Forré, P.; Pirok, B. W. J. Closed-loop automatic gradient design for liquid chromatography using Bayesian optimization. *Analytica Chimica Acta* **2023**, *1242*, 340789. DOI: https://doi.org/10.1016/j.aca.2023.340789.
(57) Stanton, S.; Maddox, W.; Gruver, N.; Maffettone, P.; Delaney, E.; Greenside, P.; Wilson, A. G. Accelerating Bayesian Optimization for Biological Sequence Design with Denoising Autoencoders. In Proceedings of the 39th International Conference on Machine Learning, Proceedings of Machine Learning Research; 2022.
(58) Ziatdinov, M. A.; Liu, Y.; Morozovska, A. N.; Eliseev, E. A.; Zhang, X.; Takeuchi, I.; Kalinin, S. V. Hypothesis Learning in Automated Experiment: Application to Combinatorial Materials Libraries. *Advanced Materials* **2022**, *34* (20), 2201345. DOI: https://doi.org/10.1002/adma.202201345.
(59) Liu, Y.; Morozovska, A. N.; Eliseev, E. A.; Kelley, K. P.; Vasudevan, R.; Ziatdinov, M.; Kalinin, S. V. Autonomous scanning probe microscopy with hypothesis learning: Exploring the physics of domain switching in ferroelectric materials. *Patterns (N Y)* **2023**, *4* (3), 100704. DOI: 10.1016/j.patter.2023.100704 From NLM.
(60) Jeong, W. H.; Yu, Z.; Gregori, L.; Yang, J.; Ha, S. R.; Jang, J. W.; Song, H.; Park, J. H.; Jung, E. D.; Song, M. H.; et al. In situ cadmium surface passivation of perovskite nanocrystals for blue LEDs. *Journal of Materials Chemistry A* **2021**, *9* (47), 26750-26757, 10.1039/D1TA08756H. DOI: 10.1039/D1TA08756H.
(61) Kulkarni, S. A.; Baikie, T.; Boix, P. P.; Yantara, N.; Mathews, N.; Mhaisalkar, S. Band-gap tuning of lead halide perovskites using a sequential deposition process. *Journal of Materials Chemistry A* **2014**, *2* (24), 9221-9225, 10.1039/C4TA00435C. DOI: 10.1039/C4TA00435C.

(62) Knight, A. J.; Borchert, J.; Oliver, R. D. J.; Patel, J. B.; Radaelli, P. G.; Snaith, H. J.; Johnston, M. B.; Herz, L. M. Halide Segregation in Mixed-Halide Perovskites: Influence of A-Site Cations. *ACS Energy Letters* **2021**, *6* (2), 799-808. DOI: 10.1021/acsenergylett.0c02475.

(63) Brennan, M. C.; Draguta, S.; Kamat, P. V.; Kuno, M. Light-Induced Anion Phase Segregation in Mixed Halide Perovskites. *ACS Energy Letters* **2018**, *3* (1), 204-213. DOI: 10.1021/acsenergylett.7b01151.

(64) Guan, M.; Li, P.; Wu, Y.; Liu, X.; Xu, S.; Zhang, J. Highly efficient green emission $Cs_4PbBr_6$ quantum dots with stable water endurance. *Opt Lett* **2022**, *47* (19), 5020-5023. DOI: 10.1364/OL.471088  From NLM PubMed-not-MEDLINE.

(65) Kerner, R. A.; Xu, Z.; Larson, B. W.; Rand, B. P. The role of halide oxidation in perovskite halide phase separation. *Joule* **2021**, *5* (9), 2273-2295. DOI: https://doi.org/10.1016/j.joule.2021.07.011.

(66) Knight, A. J.; Wright, A. D.; Patel, J. B.; McMeekin, D. P.; Snaith, H. J.; Johnston, M. B.; Herz, L. M. Electronic Traps and Phase Segregation in Lead Mixed-Halide Perovskite. *ACS Energy Letters* **2019**, *4* (1), 75-84. DOI: 10.1021/acsenergylett.8b02002.

(67) Lee, J.-W.; Kim, D.-H.; Kim, H.-S.; Seo, S.-W.; Cho, S. M.; Park, N.-G. Formamidinium and Cesium Hybridization for Photo- and Moisture-Stable Perovskite Solar Cell. *Advanced Energy Materials* **2015**, *5* (20), 1501310. DOI: https://doi.org/10.1002/aenm.201501310.

(68) Kim, Y.; Bae, C.; Jung, H. S.; Shin, H. Enhanced stability of guanidinium-based organic-inorganic hybrid lead triiodides in resistance switching. *APL Materials* **2019**, *7* (8). DOI: 10.1063/1.5109525 (acccessed 10/5/2023).

(69) Cui, D.; Yang, Z.; Yang, D.; Ren, X.; Liu, Y.; Wei, Q.; Fan, H.; Zeng, J. Color-Tuned Perovskite Films Prepared for Efficient Solar Cell Applications. *The Journal of Physical Chemistry C* **2015**, *120*. DOI: 10.1021/acs.jpcc.5b09393.

(70) Galkowski, K.; Surrente, A.; Baranowski, M.; Zhao, B.; Yang, Z.; Sadhanala, A.; Mackowski, S.; Stranks, S. D.; Plochocka, P. Excitonic Properties of Low-Band-Gap Lead–Tin Halide Perovskites. *ACS Energy Letters* **2019**, *4* (3), 615-621. DOI: 10.1021/acsenergylett.8b02243.

(71) Rajagopal, A.; Stoddard, R. J.; Hillhouse, H. W.; Jen, A. K. Y. On understanding bandgap bowing and optoelectronic quality in Pb–Sn alloy hybrid perovskites. *Journal of Materials Chemistry A* **2019**, *7* (27), 16285-16293, 10.1039/C9TA05308E. DOI: 10.1039/C9TA05308E.

(72) Hao, F.; Stoumpos, C. C.; Chang, R. P. H.; Kanatzidis, M. G. Anomalous Band Gap Behavior in Mixed Sn and Pb Perovskites Enables Broadening of Absorption Spectrum in Solar Cells. *Journal of the American Chemical Society* **2014**, *136* (22), 8094-8099. DOI: 10.1021/ja5033259.

(73) Ding, Y.; Wu, Y.; Tian, Y.; Xu, Y.; Hou, M.; Zhou, B.; Luo, J.; Hou, G.; Zhao, Y.; Zhang, X. Effects of guanidinium cations on structural, optoelectronic and photovoltaic properties of perovskites. *Journal of Energy Chemistry* **2021**, *58*, 48-54. DOI: https://doi.org/10.1016/j.jechem.2020.09.036.

(74) Jodlowski, A. D.; Yépez, A.; Luque, R.; Camacho, L.; de Miguel, G. Benign-by-Design Solventless Mechanochemical Synthesis of Three-, Two-, and One-Dimensional Hybrid Perovskites. *Angewandte Chemie International Edition* **2016**, *55* (48), 14972-14977. DOI: https://doi.org/10.1002/anie.201607397 (acccessed 2023/10/05).

(75) Sanchez, S. L.; Tang, Y.; Hu, B.; Yang, J.; Ahmadi, M. Understanding the ligand-assisted reprecipitation of $CsPbBr_3$ nanocrystals via high-throughput robotic synthesis approach. *Matter* **2023**. DOI: https://doi.org/10.1016/j.matt.2023.05.023.

(76) Higgins, K.; Ziatdinov, M.; Kalinin, S. V.; Ahmadi, M. High-Throughput Study of Antisolvents on the Stability of Multicomponent Metal Halide Perovskites through Robotics-Based Synthesis and Machine Learning Approaches. *J Am Chem Soc* **2021**, *143* (47), 19945-19955. DOI: 10.1021/jacs.1c10045  From NLM PubMed-not-MEDLINE.

(77) Jonghee Yang, B. J. L., Sergei V Kalinin, Mahshid Ahmadi. High-Throughput Automated Exploration of Phase Growth Kinetics in Quasi-2D Formamidinium Metal Halide Perovskites. *ChemRxiv* **2023**. DOI: 10.26434/chemrxiv-2023-zcvl0.

(78) Higgins, K.; Valleti, S. M.; Ziatdinov, M.; Kalinin, S. V.; Ahmadi, M. Chemical Robotics Enabled Exploration of Stability in Multicomponent Lead Halide Perovskites via Machine Learning. *ACS Energy Letters* **2020**, 3426-3436. DOI: 10.1021/acsenergylett.0c01749.

(79) Heimbrook, A.; Higgins, K.; Kalinin, S. V.; Ahmadi, M. Exploring the physics of cesium lead halide perovskite quantum dots via Bayesian inference of the photoluminescence spectra in automated experiment *Nanophotonics* **2021**. DOI: doi:10.1515/nanoph-2020-0662.

*Supplementary Information*

**Physics-driven discovery and bandgap engineering of hybrid perovskites**

Sheryl L. Sanchez,[1] Elham Foadian[1], Maxim Ziatdinov,[2,3] Jonghee Yang[1], Sergei V. Kalinin,[1] Yongtao Liu[*,2] and Mahshid Ahmadi[1*]

[1] *Institute for Advanced Materials and Manufacturing, Department of Materials Science and Engineering, The University of Tennessee Knoxville, Knoxville, Tennessee, 37996, United States*

[2] *Center for Nanophase Materials Sciences, Oak Ridge National Laboratory, Oak Ridge, TN 37831, USA*

[3] *Computer Science and Engineering Division, Oak Ridge National Laboratory, Oak Ridge, TN 37831, USA*

*Corresponding author emails:*

mahmadi3@utk.edu

liuy3@ornl.gov

**Materials**

Guanidinium Iodide (Sigma-Aldrich, ReagentPlus 99%), Methylammonium Iodide (Sigma-Aldrich, ReagentPlus 99%), Lead Iodide (Sigma-Aldrich, ReagentPlus 99%), Lead bromide (Sigma-Aldrich, 99.999% trace metals basis), Toluene (Sigma-Aldrich, anhydrous, 99.8%), N-dimethylformamide (DMF), and anhydrous dimethylsulfoxide (DMSO) are used without any further purification.

**Precursor Synthesis**

First, 0.5 M of MAI and 0.5 M of $PbI_2$ were dissolved in a 9:1 volume ratio mixture of DMF: DMSO. The precursor solution was then stirred continuously for 5 hours at room temperature to produce 0.5 M of completely dissolved $MAPbI_3$ precursor. At the same time, 0.5 M of $GAPbBr_3$ precursor solution was prepared in the same way with dissolving $PbBr_2$ and GABr in DMF: DMO. The solutions were then mixed in different volume ratios to obtain $GA_xMA_{1-x}Pb(Br_xI_{1-x})_3$.

**Thin Film Fabrication**

Thin films were cast on glass substrates which are ultrasonically cleaned in deionized water, acetone, and isopropanol respectively in an ultrasonic cleaning bath for 15 mins each. The substrates were dried by blowing dry nitrogen and treated with UV-Ozone for 20 min before spin coating the precursor solution.

The spin-coated films were fabricated using a two-step process with antisolvent treatment. The spin coating process uses 1000 rpm for 10 s followed by 3500 rpm for 25 s. 10 s prior to the end of the spin coating cycle, the antisolvent (Toluene) was added to the film. Then, the films were annealed at 100 °C for 20 min.

**Photoluminescence and absorption spectroscopy**

Photoluminescence spectroscopy was conducted utilizing the Biotek Cytation 5 Hybrid Multi-Mode Reader. The excitation wavelength was adjusted to 430 nm, while the emitted light was detected within the 500 to 850 nm range, with an increment of 1 nm. The measurements were taken 7 mm below the well plate and employed the sweep mode.

**Data Availability**

GPax: https://github.com/ziatdinovmax/gpax

Colab notebooks can be found at Github Repository: https://github.com/SLKS99/-Physics-driven-discovery-and-optimization-of-hybrid-perovskite-films/tree/main

**Data Analysis**

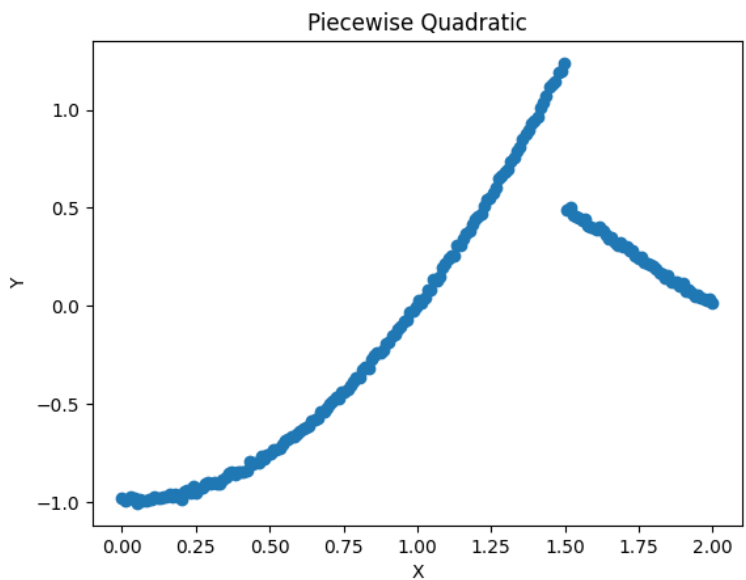

**Figure S1:** Toy model that is piecewise with a quadratic and linear interval.

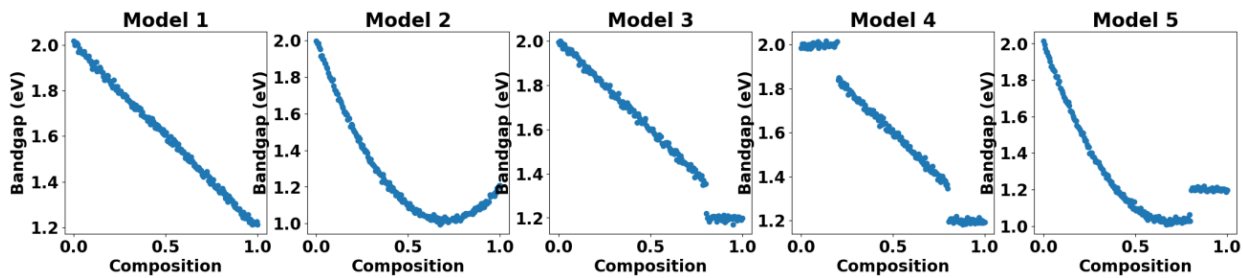

*Figure S2: Comparative Illustration of Five Bandgap Toy Models for $AB_xC_{1-X}$*

We have developed five toy models to represent the bandgap change of the mixed material with $AB_xC_{1-x}$ chemical formula, where AB has a bandgap of 1.2 eV and AC has a bandgap of 2 eV, as a function of $x$.

1. Linear Model: The first model assumes a linear relationship between the bandgap and x. The equation is given as $E_{abc} = E_{ab} \times x + E_{ac} \times (1 - x)$, where $E_{ab}$ and $E_{ac}$ represents the bandgap energies of AB and AC, respectively.

2. Bowing Model: In the second model, we introduce the concept of bandgap bowing. The equation becomes $E_{abc} = E_{ab} \times x + E_{ac} \times (1 - x) + k \times x \times (1 - x)$, where $k$ represents the bowing coefficient.

3. Alloy Formation Model with Threshold (x < 20%): The third model incorporates the condition of alloy formation. When the ratio of AC is smaller than 20%, no alloy is formed, and the bandgap is solely determined by AB. Therefore, we assume the bandgap of $AB_xC_{1-x}$ in this range is constant and equal to the bandgap of AB. The equation is given as $E_{abc} = E_{ab}$ for $x > 80\%$ and $E_{ab} \times x + E_{ac} \times (1 - x)$ for $x < 80\%$.

4. Alloy Formation Model with Thresholds ($x < 20\%$ and $x > 80\%$): The fourth model extends the alloy formation condition to include extremely small values of $x$. In this case, no alloy is formed when x is below 20% or above 80%. The equation becomes $E_{abc} = E_{ab}$ for $x \geq 80\%$, $E_{ac}$ for $x \leq 20\%$, and $E_{ab} \times x + E_{ac} \times (1 - x)$ for $20\% < x < 80\%$.

5. Alloy Formation Model with Bowing ($x < 80\%$): The fifth model combines the alloy formation condition with the bowing effect. No alloy is formed when x is greater than 80%, and the bandgap follows the bowing law when an alloy is formed. The equation is given as $E_{abc} = E_{ab}$ for $x \geq 80\%$ and $E_{ab} \times x + E_{ac} \times (1 - x) + k \times x \times (1 - x)$ for $20\% < x < 80\%$.

These toy models allow us to explore different scenarios and capture the variations in the bandgap of $AB_xC_{1-x}$ as a function of x. They provide insights into the relationship between composition and bandgap, taking into account factors such as linearity, bowing, and alloy formation conditions.

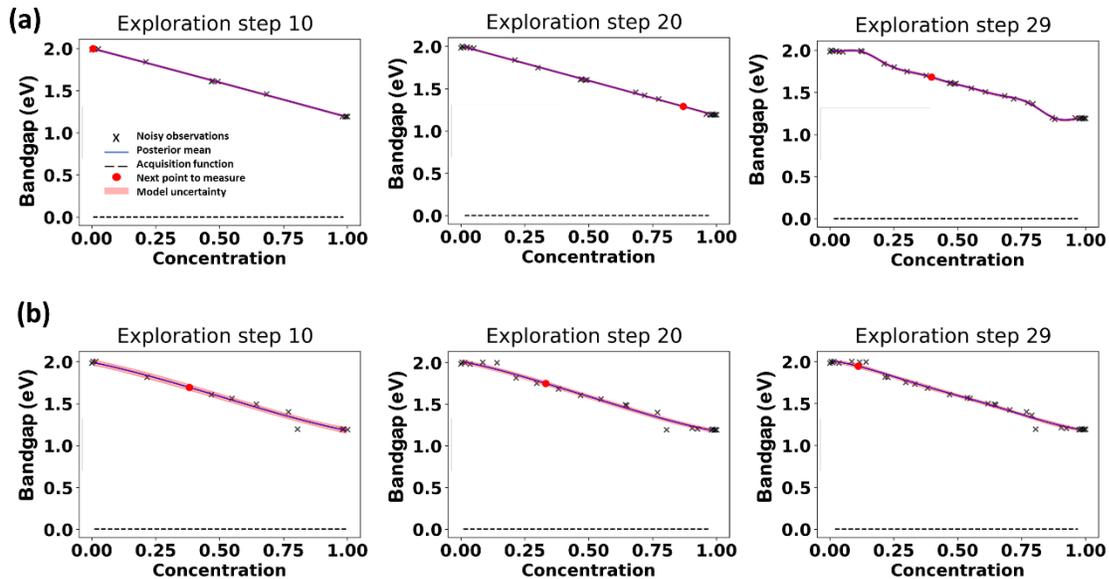

*Figure S3:* *Effect of priors on ground truth discovery in Model 4 (a) Slope prior a= Normal (0,1000), Intercept b= Normal (1000,1000) (b) Slope prior a= LogNormal (0,1), Intercept b= Normal (0,2).*

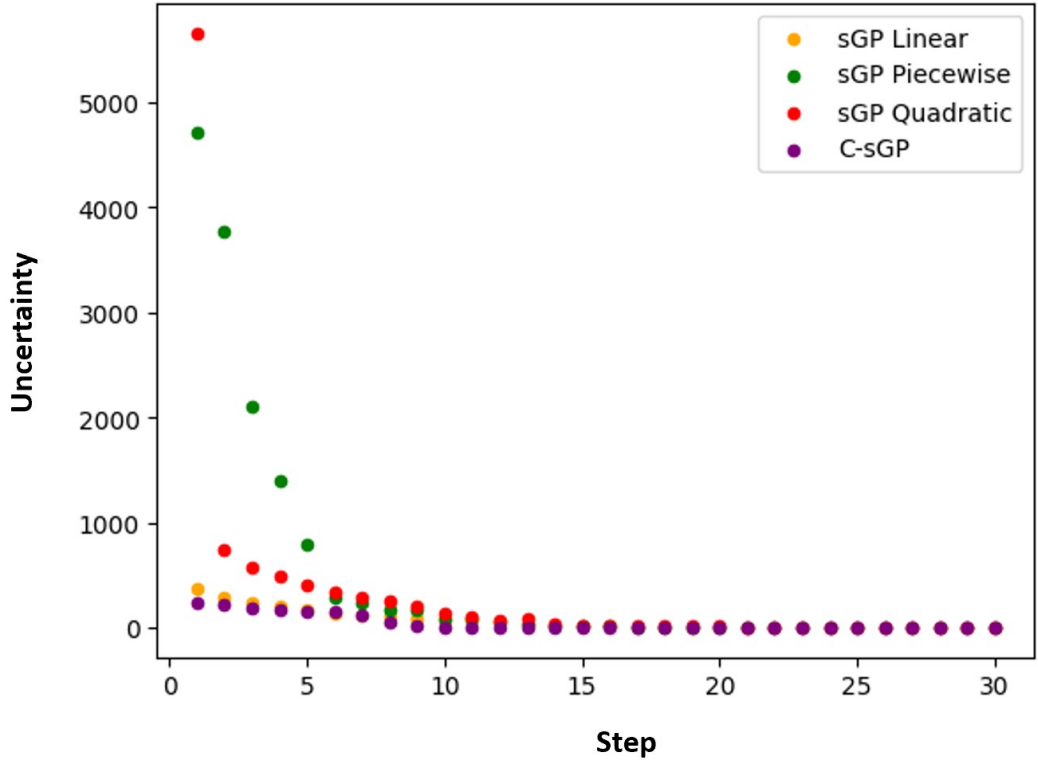

*Figure S4:* *Uncertainty d-sGP and c-sGP with mean functions linear, quadratic, piecewise and custom bandgap models.*

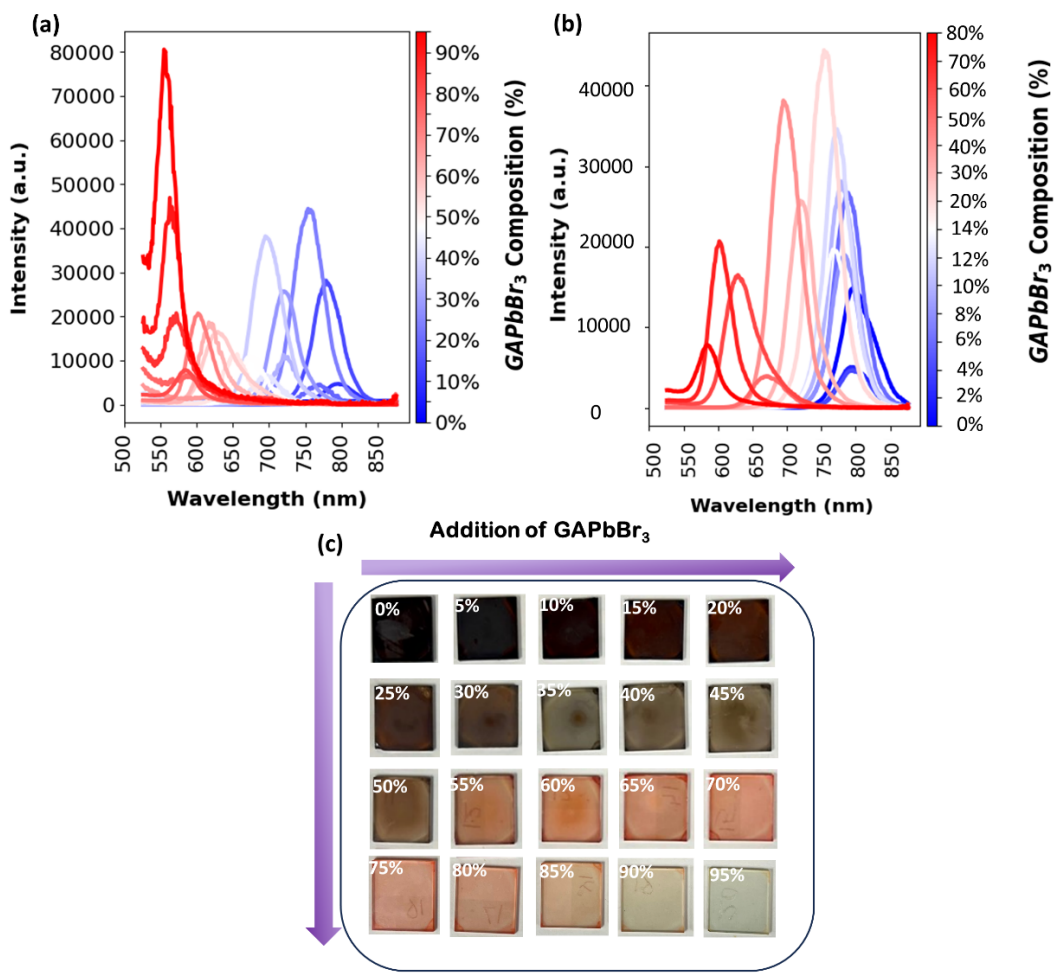

***Figure S5:*** *PL of mixed GA$_x$MA$_{1-x}$Pb(Br$_x$I$_{1-x}$)$_3$ thin films (a) with 5% increment of GAPbBr$_3$ addition (b) 2% increment up 14% then 10% increment up to 80% (c) thin films at 5% increments.*

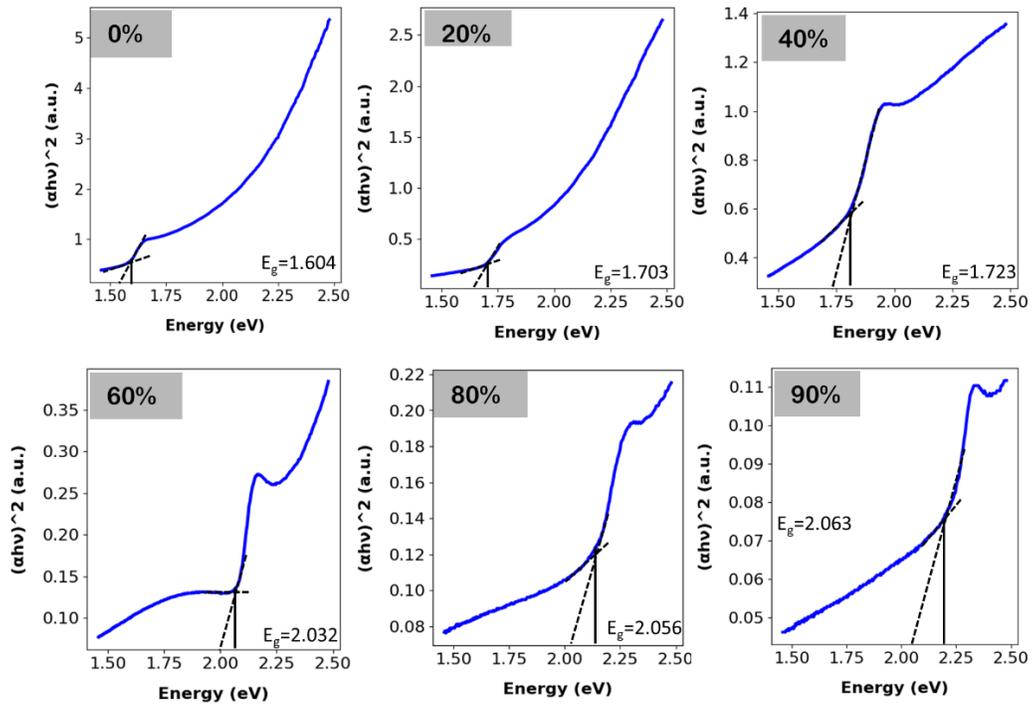

*Figure S6:* *Tauc Plot of Measured Bandgap of various addition of GAPbBr₃.*